\documentclass[%
 reprint,
 amsmath,amssymb,
 aps,
]{revtex4-2}

\usepackage{graphicx}
\usepackage{dcolumn}
\usepackage{bm}

\begin{document}

\preprint{APS/123-QED}

\title{Ion shielding effects on the resonant boundary layer response to magnetic perturbations}

\author{J. C. Waybright}
\email{jacew@princeton.edu}
\affiliation{Princeton Plasma Physics Laboratory, Princeton, New Jersey 08543, USA}%
\author{Y. Lee}%
\affiliation{%
 Department of Nuclear Engineering, Seoul National University, Seoul, South Korea
}%
\author{J.-K. Park}
\email{jkpark@snu.ac.kr}
\affiliation{%
 Department of Nuclear Engineering, Seoul National University, Seoul, South Korea}

\date{\today}

\begin{abstract}
Fusion plasmas are highly sensitive to external magnetic perturbations which result in complex responses near a region known as the resonant layer. Past analytic descriptions of this phenomena used boundary layer theory in a simplified system assuming low plasma $\beta$ to predict the onset of instabilities. Here, we present a novel extension of the analytic theory utilizing nested boundary layers to capture the physics of ion parallel flow at the plasma resonant layer. This new prediction supports previous numerical results and suggests ion shielding against magnetic disruptions in parameter regimes relevant to future device operation.
\end{abstract}

\maketitle


In the physical description of fluid flow \cite{Stagg2017, Duguet2012}, heat transfer \cite{Leiqi2023}, and particle diffusion \cite{Gaggioli2024} across real media, we observe a stark separation of scales which indicate where different physical effects are significant. The thin, inner layer covering a surface contains complex physics such as viscosity and turbulence which vary on rapid temporal and spatial scales whereas the wide outer layer behaves more ideally and varies over greater scales. We obtain a complete description by solving the corresponding fluid equations for both layers and matching the solutions in between. Since its first introduction by Prandtl in 1904 \cite{Prandtl04}, boundary layer theory has been a powerful tool for many physics disciplines including fluid dynamics, biophysics, and plasma physics \cite{Schil17,Orszag99,Verhulst05}. 

A well known boundary layer problem in fusion plasmas is the plasma response to magnetic perturbations at a rational surface. A rational surface is where the magnetic field lines are closed on themselves after $n$ poloidal and $m$ toroidal transits, and thus becomes highly vulnerable to the poloidal and toroidal $(m,n)$ mode perturbation. When plasma is sufficiently conductive as expected in fusion-grade tokamaks, the plasma response in outer layers except in the neighborhood of the rational surfaces remains ideal, with the field lines frozen to the motion of plasma. In the inner layer, however, small resistivity allows magnetic reconnection and magnetic islands, competing with other subsidiary effects including viscosity, diffusion, and flow, often in distinct ways between ions and electrons. Particularly, the problem of \textit{forced magnetic reconnection} \cite{HahmPF85}, in the context of the intrinsic error field (EF) \cite{LaHaye92, Fitzpatrick1993, Buttery99, Wolfe2005, LaHaye92, Park2007, Menard2010, PazSoldan2014} or extrinsic resonant magnetic perturbation (RMP) is of great interest for tokamaks due to its substantial impact on confinement and stability \cite{Suttrop2011, Evans2004, Kirk2012, Evans2008, Nazikian2015, Hu2019, Hu2020, Sun2016, Liang2007, Jeon2012, Ye2025}. A critical information is the onset point of the strong connection, called resonant \textit{field penetration}, where boundary layer theory becomes instrumental since the inner layer is narrow, scaling as $\sim S^{-1/3}$, and the Lundquist number, $S$, is typically $10^8\sim10^{10}$ in fusion plasmas. \\
\indent The physics of field penetration is encapsulated in the inner-layer stability index $\Delta$ \cite{FKRPF63} that represents the shielding currents against the resonant fields, which is the only information required to match the external solutions across the outer regions. Previous extensions of the theory to two-fluid and drift magnetohydrodynamics (MHD) \cite{Cole2006, Fitz2022, Park2022} have identified over 10 different parametric regimes, offering scaling predictions that align more closely with experimental observations \cite{Waybright_2024, Logan2020}. However, a persistent singularity plagues these models: the electromagnetic torque diverges ($\tau_{EM}\propto 1/\text{Im}(\Delta)$) as $\Delta\to0$, where the reconnected flux $\psi_{mn}$ as well as the size of magnetic islands $\delta_{mn}$ also increase indefinitely as $\delta_{mn}\propto\psi_{mn}\propto 1/|\Delta|$. This occurs when the $\vec{E}\times\vec{B}$ frequency approaches the electron diamagnetic frequency $\omega_{*e}$, which is called the \textit{natural mode frequency}. This singularity implies an unphysical, immediate transition to fully grown island regimes. We posit that this deficit stems from the neglect of ion parallel flow, a simplification that renders standard asymptotic matching tractable but is unjustifiable in reactor-relevant regimes. Supported by numerical evidence of altered penetration processes \cite{Lee_2024} and experimental shifts in natural frequency \cite{Koslowski_2006, DeBock_2008}, this Letter reports the first theoretical resolution of this linear singularity, via nested boundary layer matching that includes ion parallel flow. \\
\indent A notable prediction emerging from this theory is the possibility of strong resilience against field penetration. By incorporating ion parallel flow into the second Hall resistive regime (HRii) which is relevant for fusion plasmas \cite{Cole2006, Waybright_2024}, we derive the modified stability index: $\hat{\Delta}_{\text{HRii}} = \hat{\Delta}_0 + \hat{\Delta}_{\text{i}}$ with

\begin{align}
    \hat{\Delta}_0 & = 2\pi i\frac{\Gamma(\frac{3}{4})}{\Gamma(\frac{1}{4})}\bigg(\frac{c_{\beta}}{D}\bigg)^{\frac{1}{2}}(Q-Q_e), \label{eq:basedelta} \\
    \hat{\Delta}_{\text{i}} & = 2\pi \frac{\Gamma(\frac{3}{4})}{\Gamma(\frac{1}{4})}\bigg(\frac{c_{\beta}}{D}\bigg)^{\frac{3}{2}}\bigg[i^{1/2}\left((Q-Q_i)Q\right)^{\frac{1}{4}} - \frac{c_{\beta}}{D}\bigg]. \label{eq:correctiondelta}
\end{align}
Here $\hat{\Delta}_0$ is the previously predicted inner layer $\hat{\Delta}$ without ion parallel flow, and $\hat{\Delta}_{\text{i}}$ is the ion parallel flow correction term. Other parameters are $c_{\beta}= \sqrt{\beta/(1+\beta)}$, where $\beta = (5/3)P_0/B_0^2$, $P_0$ is the equilibrium plasma pressure, $B_0$ is the equilibrium toroidal magnetic field at the layer, $D$ is the normalized, stretched ion skin depth, $Q$ is the normalized, stretched $E \times B$ flow, and $Q_{i(e)}$ is the normalized, stretched ion (electron) diamagnetic flow. The frequently occurring ratio, $c_{\beta}/D$, scales with the square root of the plasma density, $n_0$. The hat symbol simply represents scaling using the stretch parameter, $\hat{\Delta} = \epsilon \Delta$, where $\epsilon = S^{-1/3}$ is the characteristic inner layer width. The correction from ion parallel flow causes the zero crossing of the imaginary part of the inner layer $\hat{\Delta}$ to shift downwards from $Q = Q_e$, and also provides a nonzero real part of $\hat{\Delta}$ removing the singularity and effectively enhancing the plasmas stability against the potentially harmful field penetration. We understand this as a shielding effect arising from the ion parallel flow generated in response to magnetic perturbations. This manifests in the $\mathbf{V}_z \times \mathbf{B}$ term in the generalized Ohm's law, representing the advection of the equilibrium field by the perturbed ion parallel flow. Similar screening effects have also been identified for $\mathbf{E} \times \mathbf{B}$ and diamagnetic flows producing a stabilizing effect against tearing modes \cite{Becoulet_2012, TANG_2023}. 

\begin{figure}
    \centering
    \includegraphics[width=1.0\linewidth]{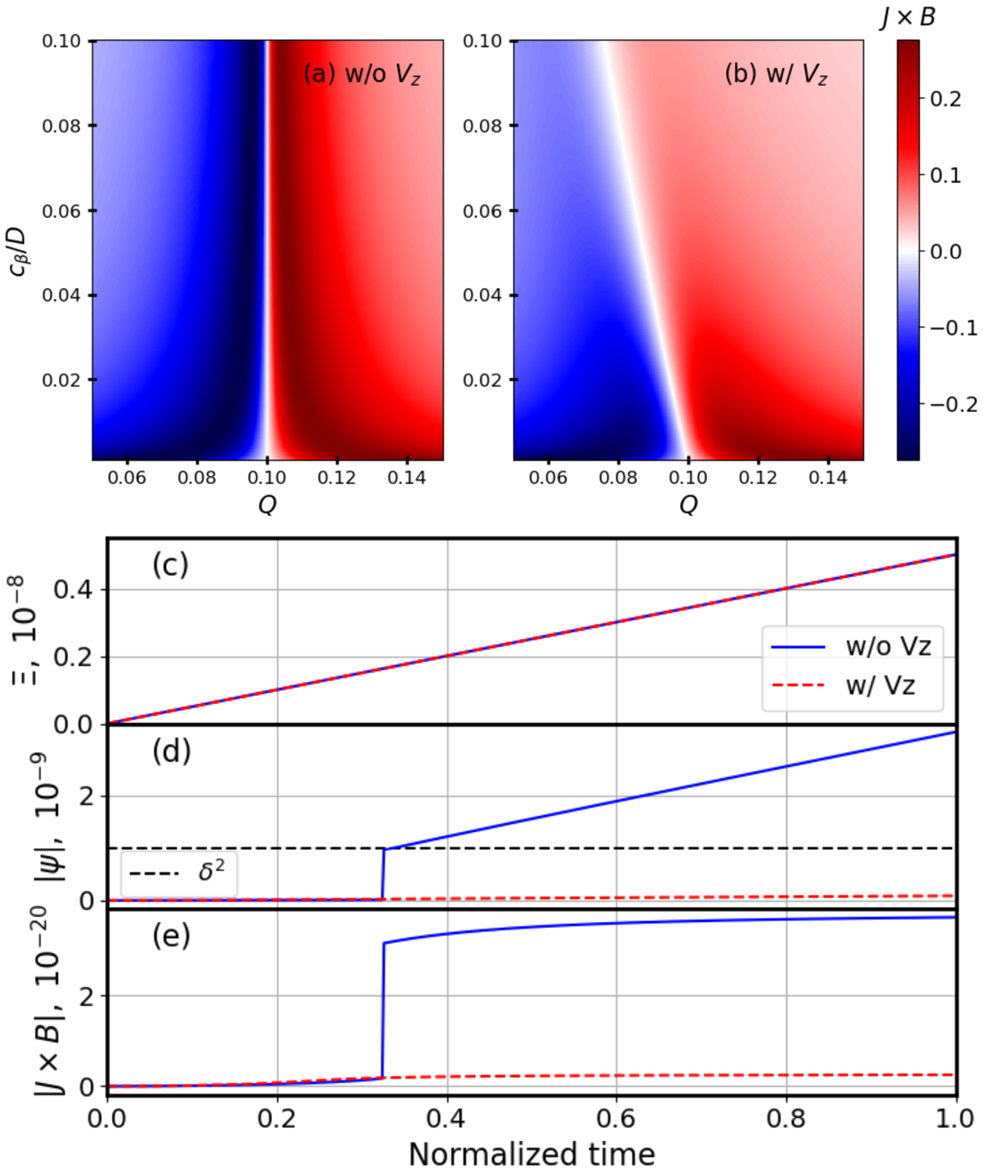}
    \caption{Magnetic braking force, $F_{J \times B}$ without ion parallel flow, $V_z$ (a) and with ion parallel flow (b) for various $c_{\beta}/D$ when $Q_e = 0.15 = -Q_i$, $\epsilon = 0.001$, $k = 1$, $\Xi=1$. The lower plot shows the time evolution of a gradually increasing error field, $\Xi$ (c), reconnected flux $|\Psi|$ (d), and magnetic braking force (e). }
    \label{fig:torque}
\end{figure}

\begin{figure}
    \centering
    \includegraphics[width=0.9\linewidth]{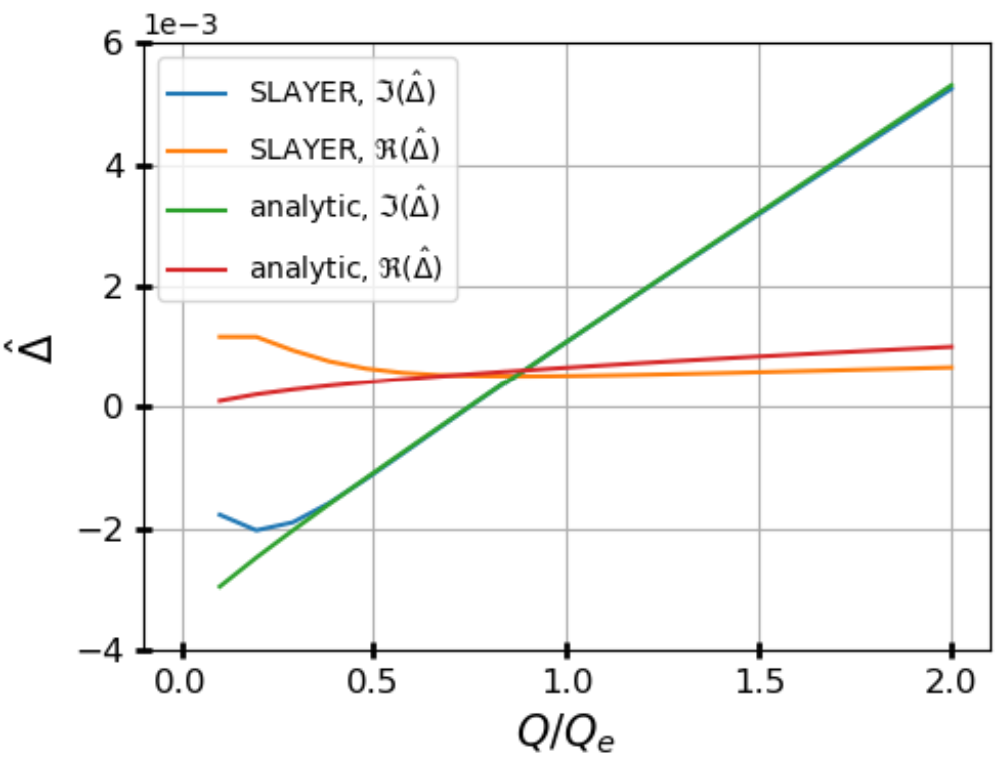}
    \caption{Comparison of numerical and analytic result for HRii regime. Here, $c_{\beta} = 0.25$, $D = 7.5$, and $Q_e=0.01$.}
    \label{fig:HRii}
\end{figure}

\begin{figure*}
    \includegraphics[width=0.9\linewidth]{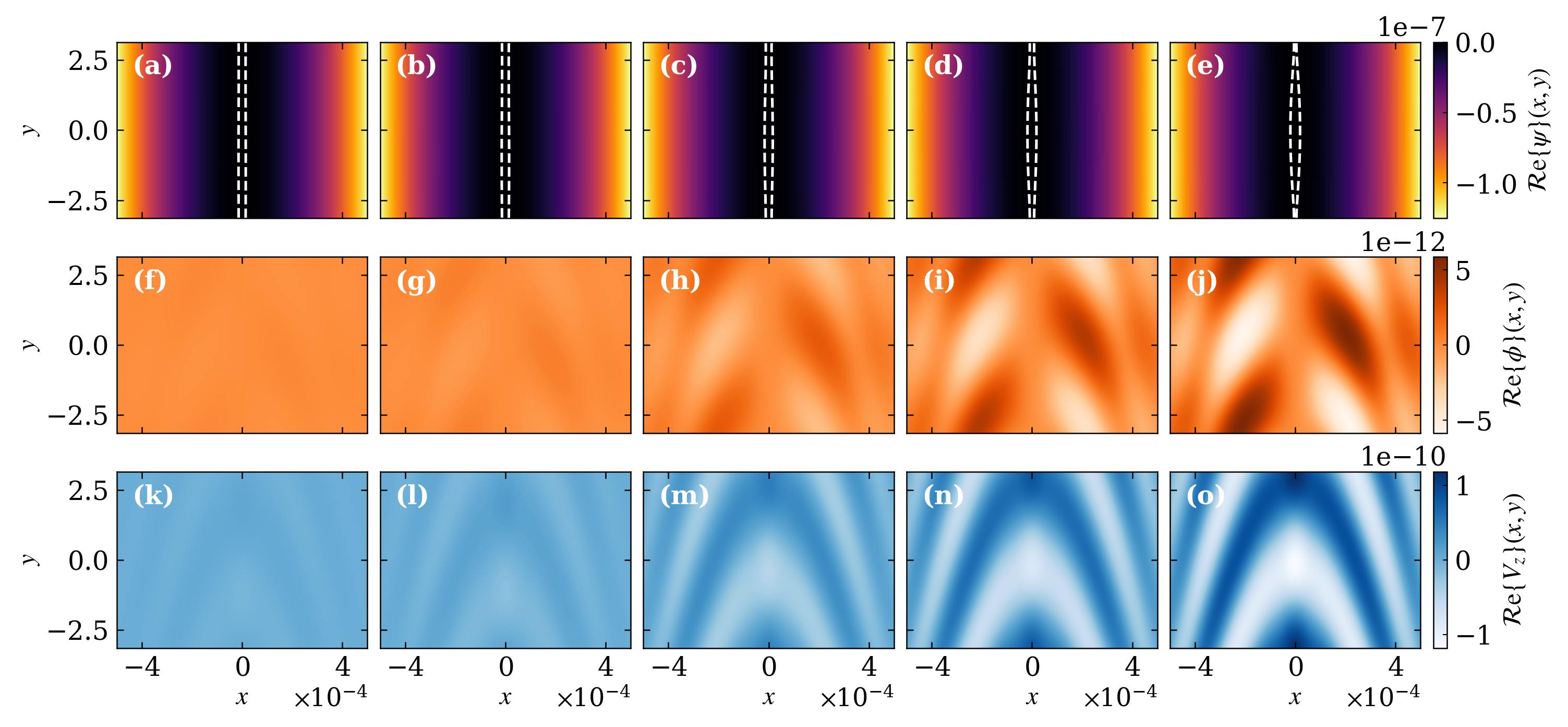}
    \caption{Evolutions of $\psi$(a-e), $\phi$(f-j) and $V_z$(k-o) during the ramp up of $\Xi$ shown in Figs. \ref{fig:torque}(c-e). Each field is reconstructed using SLAYER \cite{Lee_2024}. From the first to fifth columns, each column corresponds to normalized time = 0.1, 0.2, 0.4, 0.7 and 1.0, respectively. In (a-e), the white contour indicates the $\psi=-1\times10^{-10}$ surface.}
    \label{fig:prl_figys}
\end{figure*}

The dependencies on $c_\beta, D, Q_i$ all manifest the correction through the ion channel. It is also clear that the correction is in higher order to $c_\beta/D<1$ which is small but increases for high-performance fusion plasmas. Fig. \ref{fig:torque} shows the new prediction and the reduction of the electromagnetic torque and shift away from the electron diamagnetic frequency at the rational surface as $c_\beta/D$ increases compared to the prediction without considering ion parallel flow. This electromagnetic torque is sometimes called the magnetic braking force or $J \times B$ torque, and is calculated through
\begin{align}
    F_{J \times B} & = -\frac{k}{2}|\Psi|^2 \text{Im}(\Delta) \label{eq:brakingforce} \\
    \Psi & = \frac{\Delta_{sw}\Xi(t)}{\Delta - \Delta_{ss}}.
\end{align}
Here, $k$ is the wavenumber of the error field perturbation, $\Psi$ is the perturbed magnetic flux at the edge of the resonant layer, $\Delta_{sw} = 2k/\sinh k$ is the free energy from the resonant external field, $\Delta_{ss} = -2k/\tanh k$ is the classical tearing index, and $\Xi(t)$ is a slowly varying function representing the perturbed magnetic flux at the edge of the plasma. This reduction implies that the electromagnetic torque will eventually be in balance with
viscous torque which increases from the equilibrium rotation, with significantly smaller islands than expected from magnetic perturbations. Fig. \ref{fig:torque} also demonstrates that ion shielding enables the plasma to withstand a greater external field without penetration as indicated by the reconnected flux. It is also clear that the island width ($W \sim \sqrt{\Psi}$) remains well below the estimated linear layer width, $\delta$ which verifies the spatial validity of this treatment. The numerical verification also confirms this conclusion, as shown in Fig. \ref{fig:HRii}, the comparison with the code SLAYER (Slab Layer). SLAYER is a subroutine of GPEC which calculates the inner layer $\Delta$ using a Riccati transform and was recently improved to include the effects of ion flow \cite{Park2022, Lee_2024}. The analytic and numerical results agree remarkably well, particularly near the electron diamagnetic frequency, $Q = Q_e$, and both exhibit the two key characteristics of the ion shielding effect: a downward shift of the $\Im (\hat{\Delta})$ curve and a nonzero $\Re(\hat{\Delta})$ which removes the torque singularity. Fig. \ref{fig:prl_figys} shows the evolution of the SLAYER reconstructed perturbed magnetic flux, $\psi$, guiding center stream function, $\phi$, and ion parallel flow, $V_z$, during the incidence of the error field shown in Fig. \ref{fig:torque} demonstrating the initial opening of a magnetic island.

\begin{figure}[t!]
    \centering
    \includegraphics[width=0.9\linewidth]{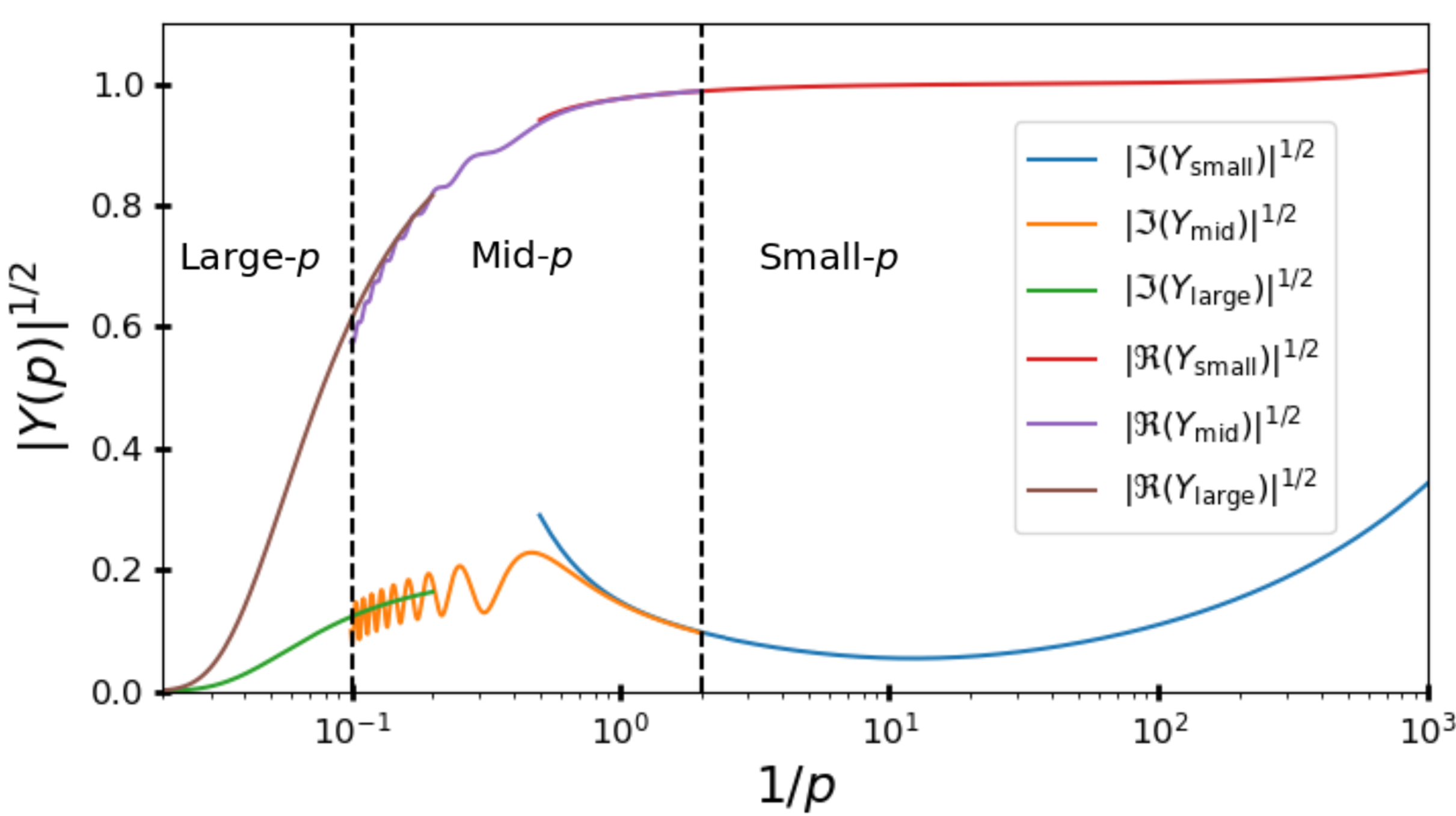}
    \caption{Full solution of real and imaginary parts of Y(p) for $c_{\beta}=0.1, D=10, Q=7.5 \times 10^{-3}, Q_e=-Q_i=0.01$ for the small, mid, and large $p$ layers}
    \label{fig:placeholderY}
\end{figure}

The construction of nested boundary layers and asymptotic matching behind this new theory is described below in detail. Consider a quasineutral plasma system with a uniform and constant ion/electron density $n_0$, and ion/electron temperatures $T_i = \tau T_e$, where $\tau$ is constant. The system is invariant in the $z$ direction, periodic in the $y$ direction, and $x$ represents the radial direction. The following is the set of normalized two fluid drift MHD equations taken from \cite{Fitzpatrick2005}, which include physics important to the resonant layer such as diamagnetic flow, finite Larmor radius, Hall effects, and phenomenological viscous and thermal diffusion,
\begin{gather}
    \mathbf{E} + \mathbf{V} \times \mathbf{B} + d_i (\mathbf{\nabla} P - \tfrac{\tau}{1 + \tau}(\mathbf{b} \cdot \mathbf{\nabla}P)\mathbf{b} \notag \\
    - \mathbf{J} \times \mathbf{B} ) = \eta \mathbf{J} \\
    (\tfrac{\partial}{\partial t} + \mathbf{V} \cdot \mathbf{\nabla} + \tfrac{\tau}{1+\tau}\mathbf{V}_* \cdot \mathbf{\nabla})\mathbf{V} - \tfrac{\tau}{1+\tau}\mathbf{V}_* \cdot \mathbf{\nabla}([\mathbf{b}\cdot \mathbf{V}]\mathbf{b}) \notag \\
    = \mathbf{J} \times \mathbf{B} - \mathbf{\nabla}P + \mu_i \mathbf{\nabla}^2 \mathbf{V}_i \\
    (\tfrac{\partial}{\partial t} + \mathbf{V} \cdot \mathbf{\nabla}) P = -\Gamma P \mathbf{\nabla} \cdot \mathbf{V} + \kappa \mathbf{\nabla}^2 P.
\end{gather} 
Here, we utilize Alfv\'enic units where all lengths are normalized by $a$, velocities are normalized by the Alfv\'en speed, $V_a$, time is normalized by $a/V_a$, and magnetic fields are normalized by scale field $B_a$. The parameter $d_i = \sqrt{m_i/n_0e^2\mu_0}/a$ is the ion skin depth, $P$ is the total plasma pressure, $\eta$ is the plasma resistivity, $\mu_{i}$ is the ion viscosity, $\Gamma=5/3$ is the ratio of specific heats, $\kappa$ is the plasma thermal conductivity, $\mathbf{V}$ is the guiding center velocity, $\mathbf{V}_i = \mathbf{V} + \mathbf{V}_* \tau/(1+\tau)$ is the ion velocity, $\mathbf{V}_e = \mathbf{V} + \mathbf{V}_* \tau/(1+\tau) - d_i\mathbf{J}$ is the electron velocity, and $\mathbf{V}_* = d_i \mathbf{b} \times \nabla P/B$ is the diamagnetic flow. We have neglected the contribution of electron viscosity and will focus on regimes where it plays a subdominant role. The ion screening plays a critical role near the tokamak plasma core with high electron density, in which the classical electron viscosity is negligible due to high electron temperature. Extension to the edge region is possible but here we will neglect it for simplicity \cite{Waybright_2024}. The notable flow effects identified in the plasma torque response originate from $\mathbf{V}_z \times \mathbf{B}$ in the generalized Ohm's law. Following a reduction process to decouple compressional Alfv\'en waves as shown by Fitzpatrick and Waelbroeck \cite{Fitzpatrick2005}, the equations can be transformed into the four field model, with the four fields being the magnetic flux function, $\psi$, the guiding center stream function, $\phi$, the perturbed $z$-directed magnetic field, $b_z$, and the parallel ion flow, $V_z$. The four fields can fully express the magnetic field and guiding center velocity as $\mathbf{B} = \nabla\psi \times \hat{\mathbf{z}} \ + (B_0 + b_z)\hat{\mathbf{z}}$, and $\mathbf{V} = \nabla \phi \times \hat{\mathbf{z}} + V_z$, where $B_0$ is the uniform, constant equilibrium field. For convenience, we will use $Z = b_z d_i/(1+\tau)$ and $\bar{V}_z = d_iV_z/(1+\tau)$, neglecting the bar moving forward. Next, we subject the system to an error field via the wall boundary condition on the magnetic flux, and incorporate equilibrium tearing-stable fields \cite{Cole2006}.  \\
\indent In the outer region far from the resonant surface, the plasma can be accurately modeled with ideal MHD. By assuming perturbations to the system are periodic in $y$ ($\tilde{\psi} \sim e^{iky - i\omega t}$ ) and imposing the error field boundary condition, we obtain the matching condition between the inner and outer layer, $\lim_{X\rightarrow \infty}\tilde{\psi} = \tilde{\psi}_0[1 + \hat{\Delta}|X|/2 + O(X^2)]$, where $X = x/\epsilon$ is the stretched variable used to describe the thin inner region with $\epsilon = S^{-1/3}$. The inner resonant layer requires significantly more physics considerations, however it is still analytically tractable. Applying the generalized Fourier transform, $\bar{\phi} = \int_C \tilde{\phi}(X)e^{ipX}dX$, simplifies the four field model to a single ordinary differential equation in $Y \equiv \bar{\phi} - \bar{Z}$ \cite{Cole2006}. The new stretched parameters for the inner region are $V_0 - \omega/k = \epsilon Q$, where $V_0$ is the $\mathbf{E} \times \mathbf{B}$ velocity, $V_{i(e)}= - \epsilon Q_{i(e)} $, $c_{\beta}d_{i}/\sqrt{1+\tau} = \epsilon D$, $\eta = \epsilon^3 k $, $\kappa = \epsilon^3 k K$ and $\mu_{i} = \epsilon^3kP$. The inner layer equation for $Y(p)$ is 
\begin{widetext}
    \begin{small}
    \begin{align}
        \underbrace{\frac{d}{dp}\bigg( \frac{p^2 Y'(p)}{i(Q-Q_e) + p^2}\bigg)}_{\mathbf{J} \times \mathbf{B}} - \underbrace{G(p)p^2 Y(p)}_{\eta \mathbf{J}, \  (\mathbf{V - V_z}) \times \mathbf{B}} - \underbrace{H(p)\frac{d}{dp}\bigg[\frac{1}{iQ + Pp^2}  \frac{d}{dp}\bigg(\frac{1}{i(Q-Q_i)p^2 + (1+\tau)Pp^4} \frac{d}{dp}\bigg( \frac{p^2 Y'(p)}{i(Q-Q_e)+p^2}\bigg) \bigg)\bigg]}_{\mathbf{V_z} \times \mathbf{B}} = 0, \label{eq:4thordersimplified}
    \end{align}
    \begin{align}
        G(p) & = \frac{-Q(Q-Q_i) + i(Q-Q_i)(P+c_{\beta}^2)p^2 + c_{\beta}^2 Pp^4}{i(Q-Q_e) + (c_{\beta}^2 + i(Q-Q_i)D^2)p^2 + (1+\tau)PD^2 p^4}, \label{eq:G(p)} \\
        H(p) & = \frac{c_{\beta}^2}{D^2}\frac{i(Q-Q_i)D^2p^2 + (1+\tau)PD^2p^4}{i(Q-Q_e) + (c_{\beta}^2 + i(Q-Q_i)D^2)p^2 + (1+\tau)PD^2p^4}. \label{eq:H(p)}
    \end{align}
    \end{small}
\end{widetext}
Each term is labeled with its corresponding origin from the two fluid drift MHD equations. In this form we have neglected several terms which are subdominant at every dominant balance in $p$-space. Specifically, these include lower order derivative terms related to $\mathbf{V}_z \times \mathbf{B}$. For simplicity, we have also ignored the contribution of thermal conductivity which is important in describing diffusion in low $\beta$ regimes, however this can easily restored by replacing $c_{\beta}^2 \rightarrow c_{\beta}^2 + (1-c_{\beta}^2)K$ within (\ref{eq:G(p)}). \cite{Fitz2022}. The matching condition to the outer layer ideal MHD solution is converted to $\lim_{p \rightarrow 0}Y(p) = Y_0(\hat{\Delta}/\pi p + 1 - Q(Q-Q_i)p^2/6)$,
and it is also required that $Y(p) \rightarrow 0$ as $p \rightarrow \infty$. The solution to Eq. \ref{eq:4thordersimplified} can be analytically estimated by defining three layers in $p$-space where unique dominant balances occur which is sometimes called a nested boundary layer or triple deck problem \cite{Neiland70, Messiter70, Liang2009}. The inclusion of parallel flow effects permits a three layer matching technique in $p$-space for some regimes as opposed to the two-layer matching technique for the three field model \cite{Cole2006, Waybright_2024, LeeWay2025}. We will denote the mid-$p$ and large-$p$ layer widths as $p_*$ and $p_{**}$, respectively. Eq. \ref{eq:4thordersimplified} is solved in each layer and the solutions are matched between layers and then to the boundary matching conditions in order to determine all necessary constants. For nearly every physical regime, a maximum of two of the three terms will enter into the dominant balance at any layer in $p$-space. Moving forward we will assume $p_*^2 \gg Q$ and $p_{**}^2 \gg Q$, the latter of which is equivalent to the constant-$\psi$ approximation \cite{Cole2006, Furth1963}. While constant-$\psi$ regimes are the primary focus of this Letter, this asymptotic method is easily generalizable to nonconstant-$\psi$ regimes. Suppose that $Q \ll c_{\beta}^2 p_{*}^2$, $Q \gg Pp_{**}^2$, and $Q \gg (c_{\beta}/D)^2$, which is referred to as the second Hall resistive regime (HRii) \cite{Cole2006, Waybright_2024}. Here, we also assume that $|Q| \sim |Q-Q_e| \sim |Q-Q_i|$. In the small-$p$ layer, when $p \sim Q^{1/2}$, the $\mathbf{V}_z \times \mathbf{B}$ effects will dominate, leading to the reduction of Eq. \ref{eq:4thordersimplified} to
\begin{align}
    \frac{d^2}{dp^2}\bigg[\frac{1}{iQ}\frac{1}{i(Q-Q_i)p^2}
    \frac{d}{dp}\bigg( \frac{p^2 Y'(p)}{i(Q-Q_e)+p^2}\bigg)\bigg] = 0 \label{eq:smallpeq},
\end{align}
which can be solved via direct integration yielding
\begin{align}
     Y_{\text{small}}(p) = a_4 + a_3\bigg(\frac{-i(Q-Q_e)}{p} + p\bigg) \notag \\ + a_2 \bigg( \frac{-(Q-Q_i)(Q-Q_e)}{6}p^2 + \frac{i(Q-Q_i)}{12}p^4\bigg) \notag \\
     + a_1 \bigg( \frac{-iQ(Q-Q_i)(Q-Q_e)}{12}p^3 - \frac{Q(Q-Q_i)}{20}p^5\bigg). \label{eq:smallY}
\end{align}
The small-$p$ limit of this solution will match to the outer layer boundary condition. Increasing $p$, the next dominant balance occurs between the $\mathbf{J} \times \mathbf{B}$ and $\mathbf{V}_z \times \mathbf{B}$ terms and has the layer width $p_* \sim (c_{\beta}/DQ)^{1/2}$. In this layer Eq. \ref{eq:4thordersimplified} reduces to 
\begin{align}
    Y''(p) + \frac{c_{\beta}^2}{D^2Q(Q-Q_i)}\frac{d^2}{dp^2}\bigg[\frac{Y''(p)}{p^2} \bigg] = 0. \label{eq:midpY}
\end{align}
This equation can be transformed to a Bessel equation in $y(p) = Y''(p)$ and then integrated twice to reveal the following solution for $Y(p)$,
\begin{gather}
    Y_{\text{mid}}(p) = b_1p^{4}(4 {}_1\tilde{F}_2 (\tfrac{3}{4}; \tfrac{3}{4}, \tfrac{7}{4}; \tfrac{\alpha^2 p^{4}}{4}) \notag \\
    -\tfrac{3}{\Gamma(\frac{7}{4})} {}_1\tilde{F}_2 (1; \tfrac{3}{4}, 2; \tfrac{\alpha^2 p^{4}}{4})) 
    + b_2p^{5}(\tfrac{5}{4 \Gamma(\frac{9}{4})} {} _1\tilde{F}_2 (1; \tfrac{5}{4}, 2; \tfrac{\alpha^2 p^{4}}{4}) \notag \\
    - {} _1\tilde{F}_2 (\tfrac{5}{4}; \tfrac{5}{4}, \tfrac{9}{4}; \tfrac{\alpha^2 p^{4}}{4})) + b_3 p + b_4, \label{eq:midpYSolution}
\end{gather}
where $_1\tilde{F}_2(a;b,c;z)$ is the regularized hypergeometric function and $\alpha^2 = -Q(Q-Q_i)D^2/4c_{\beta}^2$. In the small-$p$ limit, this solution has the form $\lim_{p \rightarrow 0} Y_{\text{mid}}(p) = b_1 ({\Gamma(\frac{3}{4})\Gamma(\frac{7}{4})})^{-1}p^{4} + b_2(4\Gamma(\frac{9}{4})\Gamma(\frac{5}{4}))^{-1}p^{5} + b_3p + b_4$, and its large-$p$ limit is $\lim_{p \rightarrow \infty}Y_{\text{mid}}(p) = b_2(-\alpha^2 \Gamma(\frac{1}{4})\Gamma(\frac{5}{4})/4)^{-1} p - 4b_1(-\alpha^2\Gamma(\frac{3}{4})\Gamma(-\frac{1}{4})/4)^{-1} + b_3 p + b_4$, when we eliminate the exponentially growing part of the solution. This elimination is necessary to obtain a physical solution and provides the condition relating $b_1$ and $b_2$, $b_2 = -b_1 (4\Gamma(\frac{5}{4})/\Gamma(\frac{3}{4}))(-\alpha^2e^{i\pi}/4)^{1/4}$ \cite{NIST:DLMF}. The final dominant balance is denoted by $p_{**} \sim (D/c_{\beta})^{1/2}$, which occurs between the $\mathbf{J} \times \mathbf{B}$ and $\eta\mathbf{J}, \ (\mathbf{V}-\mathbf{V}_z)\times \mathbf{B}$ terms and reduces Eq. \ref{eq:4thordersimplified} to
\begin{align}
    Y''(p) - \frac{c_{\beta}^2}{D^2}p^{2} Y(p) = 0. \label{eq:largepYeq}
\end{align}
The general solution is
\begin{align}
    Y_{\text{large}}(p) = c_1\mathcal{D}_{-\frac{1}{2}}(\sqrt{\frac{2c_{\beta}}{D}} \ p) + c_2 \mathcal{D}_{-\frac{1}{2}}(i\sqrt{\frac{2c_{\beta}}{D}}p),  \label{eq:largepY}
\end{align}
where $\mathcal{D}_{\nu}(x)$ is the parabolic cylinder function. When we ensure the large-$p$ limit is zero to correspond to a physical solution, the small-$p$ limit of the solution is $\lim_{p\rightarrow 0}Y_{\text{large}}(p) = c_1(1 - 2(c_{\beta}/D)^{1/2}(\Gamma(\frac{3}{4})/\Gamma(\frac{1}{4}))p)$, where the constant $c_1$ has been redefined. The final step is to connect the solutions in between the $p$ layer boundaries to solve for $\hat{\Delta}$, resulting in (\ref{eq:basedelta}) and (\ref{eq:correctiondelta}), where the latter is the correction term produced by this new matching technique. Figure \ref{fig:placeholderY} shows the $Y(p)$ solution across all layers of $p$-space demonstrating the matching in between layers at the characteristic widths $p_*$ and $p_{**}$. The oscillations observed in the large-$p$ limit of the mid-$p$ layer solution are understood as Hall MHD waves resulting from the $\mathbf{J} \times \mathbf{B}$ and $\mathbf{V}_z \times \mathbf{B}$ balance. 
This method can be applied to several parameter regimes including the Hall Resistive regimes (HRi and HRii) which have been identified as particularly susceptible to ion parallel flow effects \cite{Lee_2024}. As earlier noted, the torque singularity is removed, previously existing at the electron diamagnetic frequency, $Q_e$. This behavior exists as a pattern in other regimes, however the singularity is not always removed but rather shifted towards the ion diamagnetic frequency, $Q_i$ into what is called the drift band.

Operational plasmas span over a wide range of rotations which encompass regions of the parameter regimes presented in this Letter \cite{SPARC_2022, DiSiena_2026}. Specifically, as plasma rotation slows down ($Q$), we expect the Hall resistive regimes to be relevant to devices. In addition to rotation, operational plasma densities correspond to $c_{\beta}/D$ values up to $O(10^{-1})$ which align with our predicted regions of significant ion shielding effects. While we have not presented an exhaustive list of predictions for all regimes relevant to operational stages for fusion devices, the characteristic shift or removal of the torque singularity is consistent, suggesting the overall enhanced resistance to forced magnetic reconnection.
\\
\indent We have demonstrated the additional resilience to field penetration via ion shielding using nested boundary layers with the two fluid drift MHD model. Our analytic estimates are corroborated by numerical results with outstanding agreement and predict ion shielding to be significant for plasma parameters relevant to current and future device operation. The analytic techniques can be applied to an extensive range of plasma parameters outside of the Hall resistive regimes and opens the door to encompassing ion shielding and electron viscosity simultaneously. This offers a significant extension of the current theory surrounding the onset of field penetration in fusion plasmas.

\begin{acknowledgments}
This work was supported by the National Research Foundation of Korea (NRF) grant funded by the Korea government (MSIT) RS-2024-00350293 and RS-2023-00281276
\end{acknowledgments}

\nocite{*}

\bibliography{apssamp_v4}

\end{document}